\documentclass[twocolumn]{revtex4}[12pt]
\usepackage{graphicx}
\begin{document}
\title{Nucleus-acoustic solitons in self-gravitating magnetized quantum plasmas}
\author{$^*$D. M. S. Zaman$^{1}$, M. Amina$^{2}$, and A. A. Mamun$^{1}$}
\address{$^{1}$Department of Physics, Jahangirnagar University,
Savar, Dhaka-1342, Bangladesh\\ $^*$Email: saadzamanshaon@gmail.com} \address{$^{2}$Department of EEE,
Bangladesh University of Business and Technology (BUBT), Mirpur-2, Dhaka-1216, Bangladesh\\}
\begin{abstract}
The basic properties of the nucleus-acoustic (NA) solitary waves (SWs) are
investigated in a super-dense self-gravitating magnetized quantum plasma (SDSGMQP) system in the presence of an external magnetic field, whose constituents are the non-degenerate light (heavy) nuclei, and
non/ultra-relativistically degenerate electrons. The Korteweg-de Vries
(KdV) equation has been derived by employing the reductive perturbation
method. The NA SWs are formed with negative (positive) electrostatic (self-gravitational) potential. It is also observed that the effects of non/ultra-relativistically degenerate electron
pressure, and the obliqueness of the external magnetic field significantly change the basic properties (e.g. amplitude, width, and speed) of NA
SWs. The findings of our present analysis can be very helpful to explain the physics behind the formation of the NA SWs in astrophysical compact objects, specially neutron stars, which are briefly discussed.

\end{abstract}

\maketitle

\section{Introduction}

In classical plasmas, the plasma particle number density is relatively low. When the quantum nature of the particles of the plasmas appreciably affects its
macroscopic properties, the plasmas may be considered as quantum plasma which has a very high particle number density. It is a very emerging subfield of plasma physics.
Quantum plasmas are ubiquitous in dense astrophysical compact environments (viz. white dwarfs, neutron stars, etc.), nanowires, quantum dots, semiconductor devices,
and high gain free electron lasers \cite{Shukla2006,Shukla2008,Mamun2001,Hossen2014,R. Hossen2015,M. Hossain2012,Ema2015}.
The degenerate plasma particle number density is extremely high for neutron stars (in the order of $10^{39}~cm^{-3}$) and white dwarfs (in the order of $10^{30}~cm^{-3}$) \cite{Mamun2017a}.

Chandrasekhar investigated that the outward pressure produced by the degenerate pressure of plasma particles in astrophysical compact objects, which is neutralized
by the inward pull of their gravity \cite{Chandrasekhar1931a}. The degenerate pressure ($P$) for electron ($e$) can be expressed by Chandrasekhar as $P=K_1 n_e^{\gamma_e}$ \cite{Chandrasekhar1931a,Chandrasekhar1931b,Mamun2010a,Mamun2010b,Shukla2011,Mamun2012,Roy2012,Hossen2015},
where  $\gamma_e={5}/{3}$, and $K_e\simeq 3\Lambda_c\hbar/5$ (with $\Lambda_c=\pi\hbar/m_ec$) for non-relativistic limit
\cite{Mamun2010a,Mamun2010b,Shukla2011,Mamun2012,Roy2012,Hossen2015,Ema2015,M. R. Hossen2014,M. M. Hasan2016}, whereas  $\gamma_e={4}/{3}$, and
$K_e\simeq 3\hbar c/4$ for ultra-relativistic limit \cite{Mamun2010a,Mamun2010b,Shukla2011,Mamun2012,El-Taibany2012,Ema2015,M. R. Hossen2014,M. M. Hasan2016}. It may be included that $m_e$ is the mass of electrons, $\hbar$ is  Planck's constant divided by $2\pi$, and $c$ is the speed of light in vacuum. Recently, relativistic degenerate plasma has achieved enormous attention due to its existence in interstellar compact objects \cite{Dip2017,M. R. Hossen and L. Nahar2014,B. Hosen2016,P. R. Dip2017} and in intense laser plasma experiments \cite{Baizid2016}.

However, white dwarfs contain mainly oxygen, carbon, and helium as heavy elements as well as hydrogen
gas as light elements while neutron star contain rubidium or molybdenum nuclei as heavy elements
and helium or lithium nuclei as light elements. From the work of Garcia-Berro {\it et al.} \cite{Garcia-Berro2010}, the main constituents of compact objects (e.g. white dwarfs, neutron stars, etc.)
are degenerate electrons, and nuclei of light as
well as heavy elements. Furthermore, the concept of self-gravitational instability is a comprehensive area of research in astrophysics and plasma physics.
The asteroids, comets, planets, and stars had been formed from the gravitational collapse of astrophysical objects, such as nebulae, etc. Self-gravitational
instability between the neutral fluids had been briefly explained in Jeans prominent theory \cite{Jeans1902} on gravitational collapse of astrophysical objects
for the formation of stars, and galaxies. When internal pressure could not compete with gravitational collapse, the Jeans instability had occurred.

Plasmas with a strong magnetic field, which significantly alter trajectories of the charged particles, are said to be magnetized plasmas.
In particular, magnetized plasmas are anisotropic, responding differently to forces, which are parallel and perpendicular to the direction of
magnetic field. However, the electric field in a plasma is usually small due to high conductivity and isn't affected by Debye shielding.
If a magnetized plasma moves with a velocity $V$, the electric field ($E$) will be $E=-V\times \textbf{B}$, where $V$ is the velocity, and $\textbf{B}$ is the magnetic field.
The external magnetic field in astrophysical plasmas shows a great importance in both linear and nonlinear gravitational wave-plasma interaction and also in
the stability condition of plasma waves \cite{Goldstein2016,Marcedo1983}.

Recently, a number of authors have made a number of theoretical
investigations on the gravitational instability of nucleus acoustic (NA)
waves in degenerate quantum magnetized plasma system. Mamun and Schlickeiser \cite{A. A.
Mamun and R. Schlickeiser2016,Mamun2015} investigated the propagation of
dust-acoustic (DA) solitary (shock) waves in a weakly (strongly)
coupled opposite polarity dust-plasma medium by taking the self-gravitational potential effect. Mamun {\it et al.}
\cite{A. A. Mamun2016} studied nucleus-acoustic shock
structures in a strongly coupled self-gravitational degenerate
quantum plasma consisting of strongly-coupled non-degenerate
heavy nuclei, weakly-coupled degenerate non-relativistically light
nuclei, and non/ultra-relativistically degenerate electrons. Mamun {\it et al.} \cite{Mamun2017} have another
work on heavy nucleus acoustic (HNA) spherical solitons (SSs)
associated with HNA waves in self-gravitational degenerate
(super-dense) quantum plasmas (SGDQP). All of these studies are
limited to unmagnetized environments and the effects
of the external magnetic field on NA SWs have not been considered
in any previous work. In our present investigation, we have studied
theoretically the basic features of nucleus-acoustic (NA) solitary waves (SWs) in a super-dense self-gravitating magnetized quantum plasma (SDSGMQP) system by considering the magnetized effect.

To the best of our knowledge, no theoretical investigation has been made
to study the NA SWs in an SDSGMQP system, which contains both non-degenerate heavy and light nuclei, and both non-relativistic
and ultra-relativistic degenerate electrons. However, in our investigation, the nonlinear propagation of
nucleus-acoustic (NA) waves (where the inertia comes mainly from
the mass density of the heavy nuclei, and the restoring force is
mainly provided by the degenerate pressure of electrons) is
considered in such a realistic astrophysical plasma
system.

\section{Governing Equations}
We consider a three component magnetized quantum plasma system
consisting of non-degenerate heavy (light) nuclei, and both non-relativistic
and ultra-relativistic degenerate electrons, and demonstrate the existence of the NA SWs in such an SDSGMQP system. The dynamics of NA SWs in
the presence of the external magnetic field ($B_0$) is governed by the following equation
\begin{eqnarray}
&&\hspace*{-8mm}\partial_tn_s+ \partial_x u_s+\partial_x(n_su_s) =
0,
\label{c1}\\
&&\hspace*{-8mm}(\partial_ t+u_p\partial_x) u_p=-
\partial_x\phi-\partial_x \psi-\omega_{cp}({{u}_p}\times{\hat{z}})=0,
\label{c2}\\
&&\hspace*{-8mm}(\partial_ t+u_n\partial_x) u_n+\beta
 \partial_x\phi+\partial_x \psi -\beta \omega_{cp}({{u}_n}\times{\hat{z}})=0
\label{c3}\\
&&\hspace*{-8mm}\partial_x \phi-\frac{k_1}{n_e}\partial_x (n_e)^{\gamma},
\label{c4}\\
&&\hspace*{-8mm}\partial_x^2 \phi=(1+\mu)n_e-\mu n_n-n_p,
\label{c5}\\
&&\hspace*{-8mm}\partial_x^2
\psi=\sigma\left(n_p+\frac{\mu}{\beta}n_n\right), \label{c6}
\end{eqnarray}

where $s=n$ ($p$) for heavy (light) nuclei, $\partial_t=
\partial/\partial t$, $\partial_ x=\partial/\partial x$,
$d_t=\partial_t +u_n\partial_x$; $n_e$ ($n_s$) is the perturbed
part of electron (nuclear species) number density normalized by
its equilibrium value $n_{e0}$ ($n_{s0}$); $u_p$ is the ion fluid speed normalized by
$C_{p}=(Z_pm_ec^2/m_p)^{1/2}$ with $m_e$ ($m_p$) being the electron
(light nuclei) rest mass, $c$ is the speed of light in vacuum; $\phi$ is
the electrostatic wave potential normalized by $m_ec^2/e$ (with $e$
being the magnitude of an electron charge); $\psi$ is the
perturbed part of the self-gravitational potential normalized by
$C_p^2$; $x$ is the space variable normalized by
$\lambda_{Dp}=(m_e c^2/4\pi n_{p0}Z_pe^2)^{1/2}$; $t$ is the time
variable normalized by $\omega_{pp}^{-1}$; $\omega_{pp}=(4\pi n_{p0}e^2/mp)^{1/2}$; $\beta=Z_n m_p/Z_pm_n$
and $\mu=Z_nn_{n0}/Z_pn_{p0}$;
$\sigma=\omega_{Jp}^2/\omega_{pp}^2$ (with $\omega_{Jp}=(4\pi G
n_{p0}m_i)^{1/2}$ being the Jeans frequency for the light nuclei,
and $G$ being the universal gravitational constant); $\omega_{cp}=(B_o^2/4\pi n_{p0}m_p)^{1/2}$; $B_0$ acts along the $z$-direction, i.e.,
$\textbf{B}_0=\hat{z}B_0$ (where $\hat{z}$ is a unit vector along the $z$-direction). We have defined
$k_1=K_en_{e0}^{\gamma_e-1}/m_e c^2$.

\section{Derivation of the K-dV Equation}

We now construct a weakly nonlinear theory for the
nonlinear propagation of the NA SWs by using the reductive
perturbation method \cite{Washimi1966}. We shall adopt the stretched coordinates \cite{Baizid2016,Rimi2013} as
\begin{eqnarray}
&&\xi=\epsilon^{1/2}(l_x{\hat{x}}+l_y{\hat{y}}+l_z{\hat{z}}-V_pt),
\label{A3a}\\
&&T={\epsilon}^{3/2}t, \label{A3b}
\end{eqnarray}

where $\epsilon$ is a smallness parameter $(0 < \epsilon < 1)$
measuring the amplitude of perturbation, $V_p$ is the wave phase
velocity normalized by the NA speed ($C_p$), and $l_x$, $l_y$,
and $l_z$ are the directional cosines of the wave vector k along
the x, y, and, z axes, respectively, so that $l_x^2$ + $l_y^2$ +
$l_z^2$ = 1. It is noted here that $x$, $y$, and $z$ are all normalized by
the Debye length $\lambda_{D}$, and $T$ is normalized by the
inverse of light nuclei plasma frequency ($\omega_{pp}^{-1}$ ). We may
expand $n_s$, $u_s$, and $\phi$ in power series of $\epsilon$ as

\begin{eqnarray}
&&n_s=1+\epsilon n_s^{(1)}+\epsilon^{2}n_s^{(2)}+ \cdot \cdot
\cdot, \label{A3c}\\
&&u_{px,y}=0+\epsilon^{3/2}
u_{px,y}^{(1)}+\epsilon^{2}u_{px,y}^{(2)}+\cdot \cdot \cdot,
\label{A3d}\\
&&u_{pz}=0+\epsilon u_{pz}^{(1)}+\epsilon^{2}u_{pz}^{(2)}+\cdot
\cdot \cdot,
\label{A3e}\\
&&\phi=0+\epsilon\phi^{(1)}+\epsilon^{2}\phi^{(2)}+\cdot \cdot
\cdot, \label{A3f}
\end{eqnarray}

Now, applying Eqs. (\ref{A3a})-(\ref{A3f}) into Eqs. (\ref{c1}) -
(\ref{c6}) and taking the lowest order coefficient of $\epsilon$,
we obtain, $u_{pz}^{(1)}={l_z \phi^{(1)}}/{V_p} + {l_z \psi^{(1)}}/{V_p}$,
$n_{p}^{(1)}={l_z^2 \phi^{(1)}}/{V_p^2}+{l_z^2 \psi^{(1)}}/{V_p^2}$, $u_{nz}^{(1)}={\beta l_z \phi^{(1)}}/{V_p} + {l_z \psi^{(1)}}/{V_p}$, $n_{n}^{(1)}={\beta l_z^2 \phi^{(1)}}/{V_p^2}+{l_z^2 \psi^{(1)}}/{V_p^2}$,
$n_{e}^{(1)}=\phi^{(1)}/k_{1}$.

The linear dispersion relation can be obtained from these
equations as,
\begin{eqnarray}
&&V_p=l_z\sqrt{(\frac{\mu(1-\beta)^2 k_1}{(1+\mu)(\mu+\beta)})} \label{A3h}
\end{eqnarray}

To the lowest order of x and y-component of the momentum
equation (\ref{c2})-(\ref{c3}) we get,

\begin{eqnarray}
&&u_{py}^{(1)}=\frac{l_x}{\omega_{ci}}\frac{\mu(1-\beta)}{\mu+\beta}\frac{\partial\phi^{(1)}}{\partial\xi},
\label{A3g}\\
&&u_{px}^{(1)}=-\frac{l_y}{\omega_{ci}}\frac{\mu(1-\beta)}{\mu+\beta}\frac{\partial\phi^{(1)}}{\partial\xi},
 \label{A3h}\\
&&u_{ny}^{(1)}=\frac{l_x}{\omega_{ci}}\frac{(1-\beta)}{\mu+\beta}\frac{\partial\phi^{(1)}}{\partial\xi},
\label{A3g1}\\
&&u_{nx}^{(1)}=-\frac{l_y}{\omega_{ci}}\frac{(1-\beta)}{\mu+\beta}\frac{\partial\phi^{(1)}}{\partial\xi},
 \label{A3h1}
\end{eqnarray}

Now, applying Eqs. (\ref{A3a})-(\ref{A3h}) into (\ref{c2})-(\ref{c3}), one
can obtain from the higher order series of $\epsilon$ of the
momentum as

\begin{eqnarray}
 &&u_{py}^{(2)}=\frac{l_yV_P}{\omega_{ci}^2}\frac{\mu(1-\beta)}{\mu+\beta}\frac{\partial^2\phi^{(1)}}{\partial\xi^2},
 \label{A3i}\\
 &&u_{px}^{(2)}=\frac{l_xV_P}{\omega_{ci}^2}\frac{\mu(1-\beta)}{\mu+\beta}\frac{\partial^2\phi^{(1)}}{\partial\xi^2},
\label{A3j}\\
 &&u_{ny}^{(2)}=\frac{l_yV_P}{\omega_{ci}^2\beta}\frac{(1-\beta)}{\mu+\beta}\frac{\partial^2\phi^{(1)}}{\partial\xi^2},
 \label{A3i1}\\
 &&u_{nx}^{(2)}=\frac{l_xV_P}{\omega_{ci}^2\beta}\frac{(1-\beta)}{\mu+\beta}\frac{\partial^2\phi^{(1)}}{\partial\xi^2},
\label{A3j}
\end{eqnarray}

Using the same process, we get the next higher order continuity
equation as well as z-component of the momentum equation. Now,
combining these higher order equations together with Eqs.
(\ref{A3g})-(\ref{A3j}) and considering $\phi^{(1)}=\psi$, one can
obtain

\begin{eqnarray}
&&\frac{\partial\psi}{\partial T} + A \psi \frac{\partial
\psi}{\partial \xi}+ B \frac{\partial^3 \psi}{\partial \xi^3}= 0, \label{A3l}
\end{eqnarray}

where A and B represent the coefficients of nonlinearity, and
dispersion, respectively. These coefficients are
given by the relations
\begin{eqnarray}
&&A=\frac{P}{Q}\left[\frac{D}{k_1}+ \frac{3\mu l_z^4 G}{V_p^4F}\right]
\label{A3m}\\
&&B=\frac{P}{Q}\left[\frac{N H \beta^2-M E(\beta^2+1)}{N H \beta^2}\right.\nonumber\\
&&\hspace*{8mm}\left.+\frac{M (\beta+1)}{N \beta}-\frac{\beta^2 R}{\sigma S}\right]
\label{A3n}
\end{eqnarray}
in which
\begin{eqnarray}
&&D=(1+\mu)(\gamma-2), E=\beta(\mu+1)V_p^2, \nonumber\\
&&F=(\mu+\beta)^2, H=(\mu+\beta), \nonumber\\
&&G=(1-\beta)^3(\mu-\beta),\nonumber\\
&&M=\mu(1-l_z^2)(1-\beta), \nonumber\\
&&N=\omega_ci^2(\mu+\beta), P=V_p^3(\mu+\beta) \nonumber\\
&&S=(\mu+\beta)^2, R=(1+\mu)^2, \nonumber\\
&&Q=2\mu l_z^2(1-\beta)^2.\nonumber
\end{eqnarray}

Equation (\ref{A3l}) represents the K-dV equation which
describes the nonlinear evolution of the NASWs in magnetized plasma.

\begin{figure}[t!]
\centering
\includegraphics[width=5.2cm]{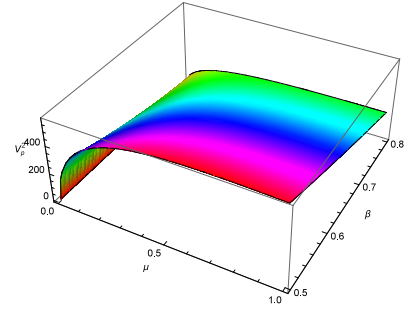}

\Large{(a)}\vspace{1cm}

\includegraphics[width=5.2cm]{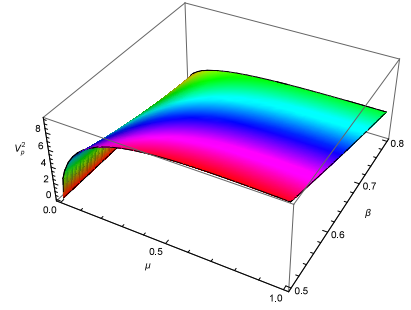}

\Large{(b)} \caption{Showing the variation of square of the phase speed ($V_p^2$) with the heavy to light nuclei density
ratio ($\mu$), and ratio of the heavy to light nuclei mass density ($\beta$) (a) for the non-relativistic case and (b) for the ultra-relativistic case.}
 \label{Fig1}
\end{figure}

\section{Numerical Analysis}

We note here that the self-gravitational field has created impact in both nonlinear
and dispersion coefficients, $A$ and $B$, via
$\mu\beta$. The steady-state solution of this K-dV equation can be obtained by choosing a moving frame $\zeta=\xi-U_0T$ (where $\zeta$ is normalized by
$\lambda_{Dp}$ and the moving frame speed $U_0$  is normalized by
$C_p$), and by imposing the appropriate boundary conditions, viz.
$\psi\rightarrow 0$, $d\psi/d\zeta\rightarrow 0$,
${d^{2}\psi}/{d\zeta^{2}}\rightarrow 0$ at ${\zeta \rightarrow
\pm\infty}$.
Therefore, the solitary wave solution is
then given by,
\begin{eqnarray}
\psi=\psi_m sech^{2}\left(\frac{\zeta}{\Delta}\right),
\end{eqnarray}

where $\psi_m$ (height normalized by $m_ec^2/e$), and
$\Delta$ (thickness normalized by $\lambda_{Dp}$) are given by
\begin{eqnarray}
\psi_m=\frac{3U_0}{A}~~{\rm and}~~\Delta=\sqrt{\frac{4B}{U_0}}.
\label{am-wid}
\end{eqnarray}

\begin{figure}[t!]
\centering
\includegraphics[width=5.2cm]{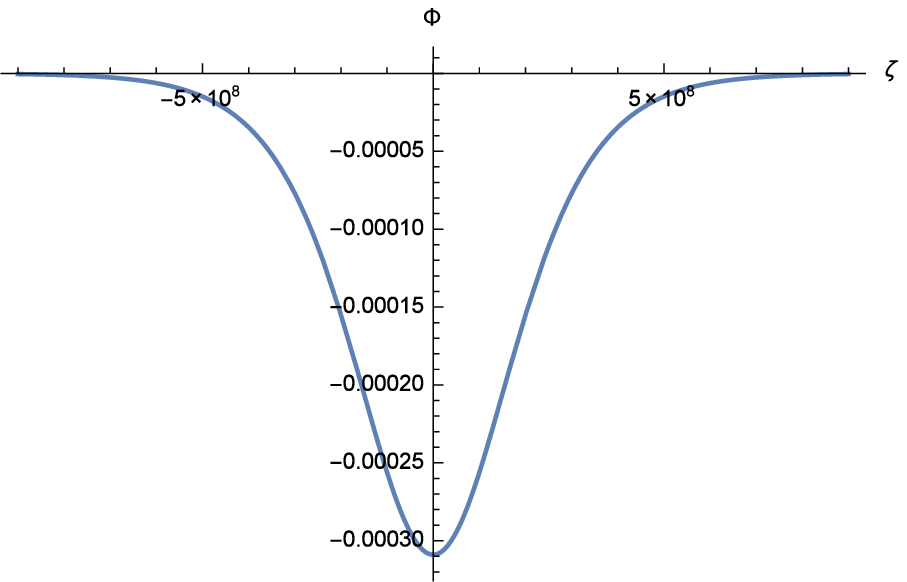}%\hspace{0.5cm}

\Large{(a)}\vspace{1cm}

\includegraphics[width=5.2cm]{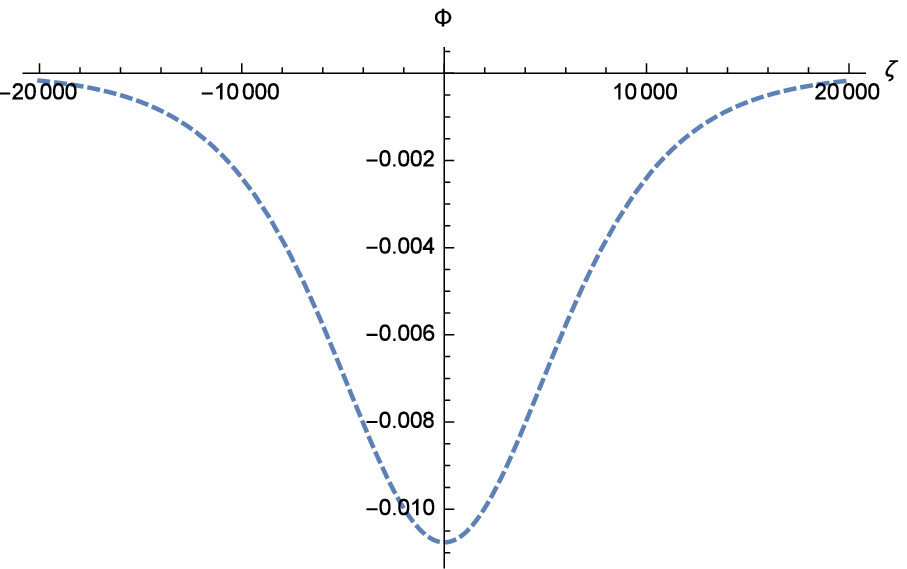}

\Large{(b)} \caption{The NA SWs associated with the electrostatic potential profiles (a) for
non-relativistic case and (b) for ultra-relativistic case.}
 \label{Fig2}
\end{figure}

By using the expressions for nonlinearity and dispersion co-efficients $A$ and $B$, we may get some
numerical appreciations of our results, viz. the solitary
wave height and width can be numerically analyzed.

The solitary waves are formed due to the balance between nonlinearity and dispersion. From equation (\ref{am-wid}), it is seen that as $U_0$ increases, the amplitude (width) of the SWs increases (decreases). It is observed
from the equation (\ref{am-wid}) that the height of the amplitude of the solitary structures is directly proportional
to the solitary speed moving with $U_0$, and inversely proportional to the nonlinear coefficient $A$. On the other
hand, the width of these solitary structures is directly proportional to the dispersive constant $B$, and inversely
proportional to the solitary speed moving with $U_0$.

\begin{figure}[t!]
\centering
\includegraphics[width=5.2cm]{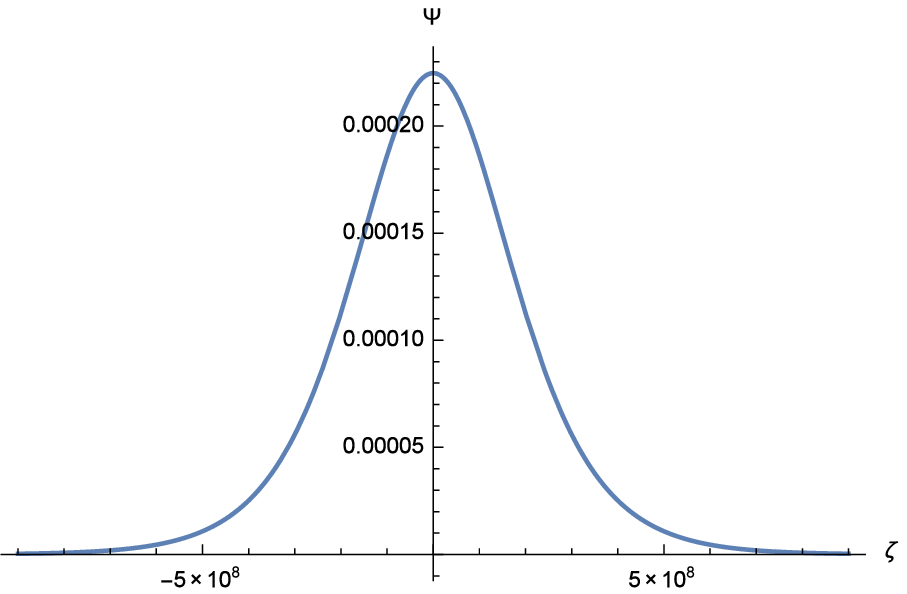}

\Large{(a)}\vspace{1cm}

\includegraphics[width=5.2cm]{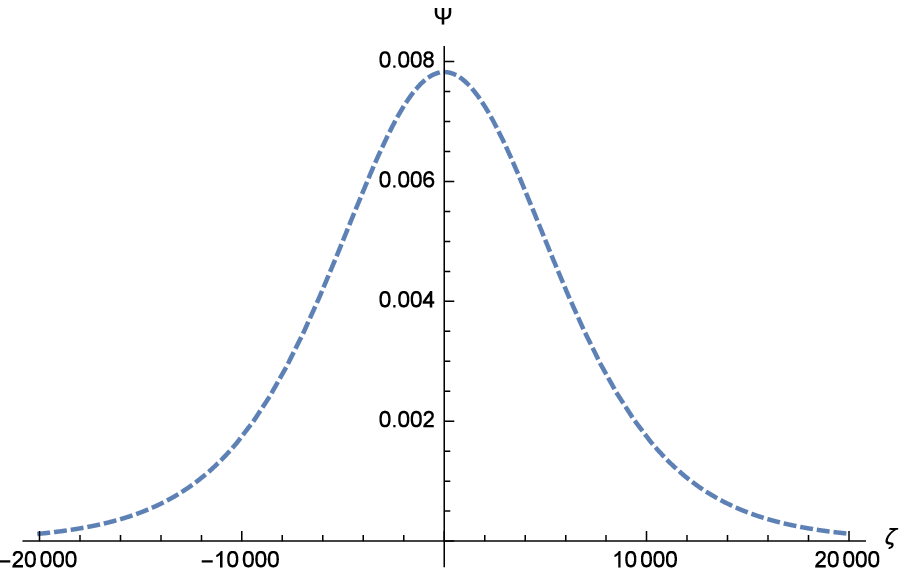}

\Large{(b)} \caption{The NA SWs associated with the self-gravitational potential profiles (a) for
non-relativistic case and (b) for ultra-relativistic case.}
 \label{Fig3}
\end{figure}

Now we have first numerical anlysed $V_p^2$ (square of the
phase speed), as a function of $\beta=Z_n m_p/Z_pm_n$
to find the parametric regimes for the formation the NA SWs with either positive or negative potential. The results are
depicted in figure \ref{Fig1} for both non/ultra-relativistic case. With the increase of $\mu$, the value of $V_p^2$ increases immediately, and after a certain value, the value of $V_p^2$ remains constant. Also the value of $V_p^2$ decreases with the increase of $\beta$ in the both non/ultra-relativistic case. It is also observed that the phase speed $V_p$ is always greater
for non-relativistic case than ultra-relativistic case. In figure \ref{Fig2}, the NA SWs associated with electrostatic potential profiles are illustrated for both non/ultra-relativistic case. It is noticed that the amplitude of NA SWs is always greater in the ultra-relativistic case than in the non-relativistic case due to high plasma particle number density. The NA SWs associated with the self-gravitational potential profiles are illustrated in figure \ref{Fig3} for both non/ultra-relativistic case. It is noted that the self-gravitational potential of NA SWs is always greater for ultra-relativistic case than non-relativistic case due to high plasma particle number density.
For NA SWs in strongly-coupled SDSGMQP,
we have considered both non/ultra-relativistically degenerate electron species (DES), and hydrogen nuclei ($H$:
$Z_i=1$, and $m_i=m_p$ where $m_p$ is the proton mass) as
degenerate light nuclear species, while carbon nuclei ($c$:
$Z_n=6$ and $m_n \simeq 12m_p$) as non-relativistic heavy nuclear
species so that all the parameters except $\mu$ are fixed. We have used $\beta=0.5$, $\mu=0.6$, $w=0.5$, $n_e=3\times10^{40}~cm^{-3}$, $n_p=2\times10^{40}~cm^{-3}$, $n_n=1.1\times10^{39}~cm^{-3}$, and $U_0=0.01$ in figures \ref{Fig2} and
\ref{Fig3}. From figures \ref{Fig2} and \ref{Fig3}, it can be concluded that the effect of ultra-relativistic DES elevates the magnitude
of the amplitude of the NA SWs associated
with both electrostatic and self-gravitational potential.

\begin{figure}[t!]
\centering
\includegraphics[width=5.2cm]{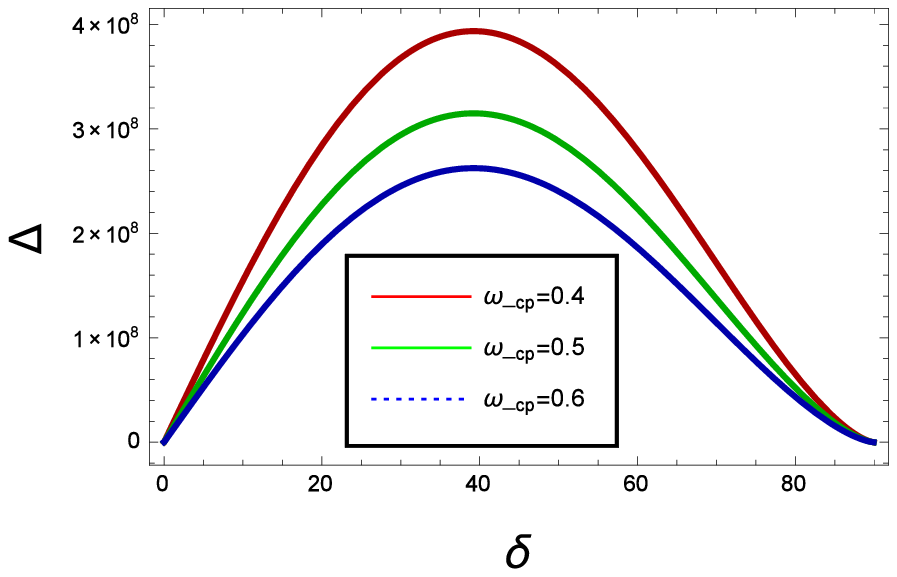}

\Large{(a)}\vspace{1cm}

\includegraphics[width=5.2cm]{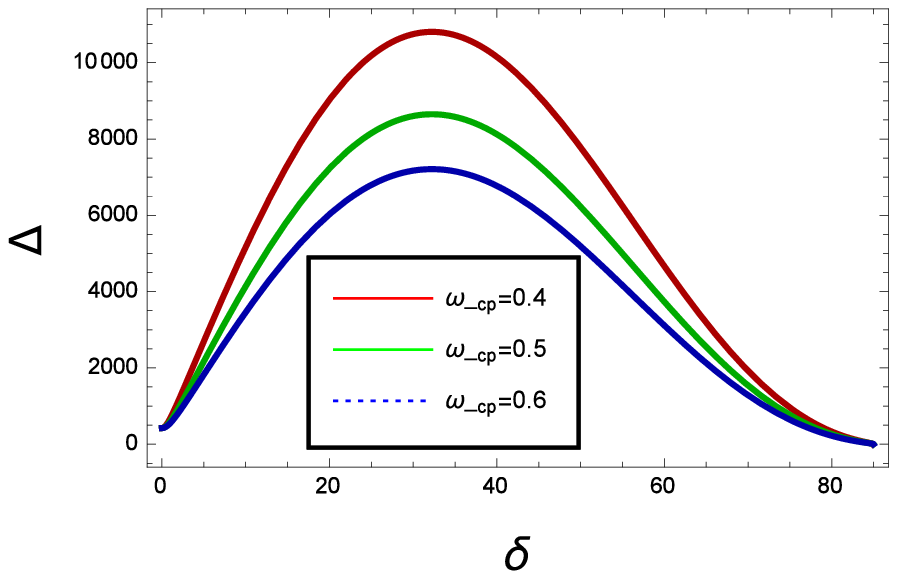}

\Large{(b)}\caption{Showing the variation of width  of the NA SWs
($\Delta$) with $\delta$ and $\omega_{ci}$ for (a) the non-relativistic and (b) for the
ultra-relativistic case.}
 \label{Fig4}
\end{figure}

The external magnetic field (${B_0}$) plays an important role on the
profile of the NA SWs. The effects of obliqueness ($\delta$) and the external magnetic
field (${B_0}$) (viz., $\omega_{ci}$) create an impact on the width of the solitary
excitations ($\Delta$) of the NA SWs for (a) non-relativistic and (b)
ultra-relativistic case, which has been depicted in figure \ref{Fig4}. It is found that the
width of the solitary profile for ultra-relativistic case is very
much smaller than the non-relativistic case.

\section{Discussion}

We have considered a magnetized plasma system (containing non-degenerate heavy (light) nuclei, and both non-relativistic and ultra-relativistic
degenerate electrons, and studied NA waves by deriving K-dV equation. We have used the reductive perturbation method to derive the partial
differential equation and found solitary wave solutions. We observed that the
relativistic effect and degenerate pressure have a great
contribution on amplitude, phase velocity, and width of the NA SWs. The results, which have been found from this
theoretical investigation, can be summarized as follows:

\begin{itemize}
  \item {The value of the square of the phase speed ($V_p^2$) increases abruptly for $\mu<0.3$ for both non/ultra-relativistic case. When $\mu\geq0.3$, the value of the square of the phase speed ($V_p^2$) remains steady for both non/ultra-relativistic case. But the value of the square of the phase speed ($V_p^2$) decreases linearly with $\beta$ for both non/ultra-relativistic case.}

  \item {The ultra-relativistically DES reduce the NA phase speed ($V_p$). It is happened because the restoring force, which is provided by ultra-relativistically DES, is less than that provided by non-relativistically DES because of the lower values of both $\gamma_e$  and $K_e$ in the ultra-relativistic limit.}

  \item {The NA SWs with negative (positive) electrostatic (self-gravitational) potential is formed because of the strong interaction among heavy nuclei which acts as source of dispersion.}

  \item {The magnitude of the amplitude and the width of the NA SWs associated with electrostatic (self-gravitational) potentials are greater in the case of ultra-relativistically DES than in the case of non-relativistically
        DES. The ultra-relativistically DES elevate the amplitude of the NA SWs due to high plasma particle number density in super-dense plasma.}

  \item {It is seen
        that in case of the non-relativistic case, the SWs width ($\Delta$) increases almost linearly for
        the lower range of $\delta$ (from $0^0$ to about $40^0$) and
        above this range $\Delta$ decreases with the increase of
        $\delta$. The width goes to zero for $\delta\longrightarrow90^0$,
        thus the amplitude leads to $\infty$. Also, the applied magnetic
        field $B_0$ plays a vital role on the width of the solitary
        profile. With the increase of the value of $\omega_{cp}$, the width of
        the solitary profile decreases.
        Thus the external magnetic field makes the
        solitary profile more spiky.}

   \item {In case of the ultra-relativistic case, the SWs width ($\Delta$) increases almost linearly for
        the lower range of $\delta$ (from $0^0$ to about $33^0$) and
        above this range $\Delta$ decreases with the increase of
        $\delta$. The width goes to zero for $\delta\longrightarrow90^0$,
        thus the amplitude leads to $\infty$. Also, the applied magnetic
        field $B_0$ plays a vital role on the width of the solitary
        profile. With the increase of the value of $\omega_{cp}$, the width of
        the solitary profile decreases. Thus a more spiky solitary profile is obtained by the effect of external magnetic field.}

   \item {The effect of ultra-relativistically DES reduce the width of the solitary
        profile in comparison to non-relativistically DES, and create a more spiky solitary profile.}
\end{itemize}

In our theoretical analysis, we have shown the
influence of obliqueness, external magnetic field, and other plasma parameters on the basic features
(phase speed, amplitude, width etc.) of the NA SWs, which
makes our present work significant in many space and astrophysical plasma
environments (viz. white dwarfs, neutron stars, and black holes).
We have observed that the nonlinear wave properties for
non-relativistic case are extremely different from
ultra-relativistic case. The astrophysical compact objects, like black holes, emit gravitational waves \cite{Abbott2016} and this recent discovery leads us to dream that, in near future, an identical or slightly identical waves (like NA waves) and associated
nonlinear structures like solitons, shocks etc., will be discovered from other astrophysical compact objects like white dwarfs and neutron stars, compact planets
like massive Jupiter, other exotic dense stars, and black holes. We may hope that this investigation will directly or indirectly be able to verify the future signature of the existence of solitonic
signals observed by space experiments as well as to understand the localized electrostatic or gravitational disturbances  in
astrophysical compact objects like white dwarfs and neutron stars.


\begin{references}

\bibitem{Shukla2006} P. K. Shukla and B. Eliasson, Phys. Rev. Lett. {\bf96}, 245001 (2006).

\bibitem{Shukla2008} P. K. Shukla, J. Plasma Phys. {\bf74}, 107 (2008).

\bibitem{Mamun2001} A. A. Mamun and P. K. Shukla, Phys. Lett. A {\bf290}, 173 (2001).

\bibitem{Hossen2014} M. A. Hossen, M. R. Hossen, and A. A. Mamun, J. Korean Phys. Soc. {\bf65}, 1883 (2014).

\bibitem{R. Hossen2015} M. R. Hossen and A. A. Mamun, Braz. J. Phys.  {\bf45}, 200 (2015).

\bibitem{M. Hossain2012} M. M. Hossain and A. A. Mamun, J. Phys. A: Math. Theor. {\bf45}, 125501 (2012).

\bibitem{Ema2015} S. A. Ema, M. R. Hossen, and A. A. Mamun, Contrib. Plasma Phys. {\bf55}, 551 (2015).

\bibitem{Mamun2017a} A. A. Mamun, Phys. Plasmas. {\bf24}, 102306 (2017).

\bibitem{Chandrasekhar1931a} S. Chandrasekhar, Phi. Mag. {\bf11}, 592 (1931).

\bibitem{Chandrasekhar1931b} S. Chandrasekhar, Astrophys. J. {\bf74}, 81 (1931).

\bibitem{Koester2002} D. Koester, Astron. Astrophys. {\bf11}, 33 (2002).

\bibitem{Mamun2010a} A. A. Mamun and P. K. Shukla, Phys. Plasmas. {\bf17}, 104504 (2010).

\bibitem{Mamun2010b} A. A. Mamun and P. K. Shukla, Phys. Lett. A {\bf374}, 4238 (2010).

\bibitem{Shukla2011} P. K. Shukla, A. A. Mamun, and D. A. Mendis, Phys. Rev.  E {\bf84}, 026405 (2011).

\bibitem{Mamun2012} A. A. Mamun, P. K. Shukla, and D. A. Mendis, J. Plasma Phys. {\bf78}, 143 (2012).

\bibitem{Roy2012} N. Roy, Phys. Plasmas {\bf19}, 064704 (2012).

\bibitem{Hossen2015} M. R. Hossen, M. A. Hossen, S. Sultana, and A. A. Mamun, Astrophys. Space Sci. {\bf357}, 34 (2015).

\bibitem{M. R. Hossen2014} M. R. Hossen, L. Nahar, and A. A. Mamun, Phys. Scr. {\bf89}, 105603 (2014).

\bibitem{M. M. Hasan2016} M. M. Hasan, M. A. Hossen, A. Rafat, and A. A. Mamun, Chin. Phys. B. {\bf25}, 105203 (2016).

\bibitem{El-Taibany2012} W. F. El-Taibany and A. A. Mamun, Phys. Rev. E {\bf85}, 026406 (2012).

\bibitem{Dip2017} P. R. Dip, M. A. Hossen, M. Salahuddin, and A. A. Mamun, Eur. Phys. J. D {\bf71}, 52 (2017).

\bibitem{M. R. Hossen and L. Nahar2014} M. R. Hossen, L. Nahar, S. Sultana, and A. A. Mamun, High Energ. Density Phys. {\bf13}, 13 (2014).

\bibitem{B. Hosen2016} B. Hosen, M. G. Shah, M. R. Hossen, and A. A. Mamun, Eur. Phys. J. Plus {\bf131}, 81 (2016).

\bibitem{P. R. Dip2017} P. R. Dip, M. A. Hossen, M. Salahuddin, and A. A. Mamun, J. Korean Phys. Soc. {\bf70}, 777 (2017).

\bibitem{Baizid2016} B. Hosen, M. Amina, A. A. Mamun, and M. R. Hossen, J. Korean Phys. Soc. {\bf69}, 1762 (2016).

\bibitem{Garcia-Berro2010}  E. Garcia-Berro, S. Torres, L. G. Althaus, I. Renedo, P. Lore´n-Aguiltar, A. H. Co´rsico, R. D. Rohrmann, M. Salaris, and J. Isern, Nature  (London)  {\bf 465}, 194  (2010).

\bibitem{Jeans1902} J. H. Jeans, Philos. Trans. R. Soc. Lond. A {\bf199}, 1 (1902).

\bibitem{Goldstein2016} P. Goldstein and E. Infeld, J. Plasma Phys. {\bf82}, 905820603 (2016).

\bibitem{Marcedo1983} P. G. Marcedo and A. H. Nelson, Phys. Rev. D {\bf28}, 10 (1983).

\bibitem{A. A. Mamun and R. Schlickeiser2016} A. A. Mamun and R. Schlickeiser, Phys. Plasmas. {\bf23}, 034502 (2016).

\bibitem{Mamun2015} A. A. Mamun and R. Schlickeiser, Phys. Plasmas.  {\bf22}, 103702 (2015).

\bibitem{A. A. Mamun2016} A. A. Mamun, M. Amina, and R. Schlickeiser, Phys. Plasmas. {\bf23}, 094503 (2016).

\bibitem{Mamun2017} A. A. Mamun, M. Amina, and R. Schlickeiser, Phys. Plasmas. {\bf24}, 042307 (2017).

\bibitem{Washimi1966} H. Washimi and T. Taniuti, Phys. Rev. Lett. {\bf 17}, 996 (1966).

\bibitem{Rimi2013} T. Akhter, M. M. Hossain and A. A. Mamun, African Phys. Rev. {\bf8}, 0012 (2013).

\bibitem {Abbott2016} B. P. Abbott, R. Abbott, T. D. Abbott, M. R. Abernathy, F. Acernese, K. Ackley, C. Adams, T. Adams, P. Addesso, R. X. Adhikari, V. B. Adya, C. Affeldt, M. Agathos, K. Agatsuma, N. Aggarwal, O. D. Aguiar {\it et al.} Phys. Rev. Lett. {\bf116}, 061102 (2016).

\end{references}
\end{document}